\documentclass[%
 reprint,
 superscriptaddress,
 amsmath,amssymb,
 aps,esint
]{revtex4-2}

\usepackage{graphicx}
\usepackage{dcolumn}
\usepackage{bm}
\usepackage{hyperref}
\usepackage{placeins}
\hypersetup{
    colorlinks=true,
    allcolors=blue,
    pdftitle={Overleaf Example},
    pdfpagemode=FullScreen,
    }

\begin{document}

\preprint{APS/123-QED}

\title{Ising Machine Based on Electrically Coupled Spin Hall Nano-Oscillators}

\author{Brooke C. McGoldrick}
\email{bcmcgold@mit.edu}
\affiliation{Department of Electrical Engineering and Computer Science, Massachusetts Institute of Technology, Cambridge, Massachusetts 02139, USA}
\author{Jonathan Z. Sun}
\affiliation{IBM T. J. Watson Research Center, Yorktown Heights, New York 10598, USA}
\author{Luqiao Liu}
\affiliation{Department of Electrical Engineering and Computer Science, Massachusetts Institute of Technology, Cambridge, Massachusetts 02139, USA}

\date{\today}

\begin{abstract}
The Ising machine is an unconventional computing architecture that can be used to solve NP-hard combinatorial optimization problems more efficiently than traditional von Neumann architectures. Fast, compact oscillator networks which provide programmable connectivities among arbitrary pairs of nodes are highly desirable for the development of practical oscillator-based Ising machines. Here we propose using an electrically coupled array of GHz spin Hall nano-oscillators to realize such a network. By developing a general analytical framework that describes injection locking of spin Hall oscillators with large precession angles, we explicitly show the mapping between the coupled oscillators' properties and the Ising model. We integrate our analytical model into a versatile Verilog-A device that can emulate the coupled dynamics of spin Hall oscillators in circuit simulators. With this abstract model, we analyze the performance of the spin Hall oscillator network at the circuit level using conventional electronic components and considering phase noise and scalability. Our results provide design insights and analysis tools toward the realization of a CMOS-integrated spin Hall oscillator Ising machine operating with a high degree of time, space, and energy efficiency.
\end{abstract}

\maketitle


\section{\label{sec:Introduction}Introduction}

As Moore's Law scaling comes to an end, new devices and architectures are becoming necessary to enable further improvements in computing efficiency. One such alternative computing paradigm is the Ising machine, which has been proposed as an efficient combinatorial optimization problem solver. Combinatorial optimization problems are ubiquitous in real-world applications including artificial intelligence, VLSI circuit design, computer networking, and industries such as steel-cutting \cite{duHandbookCombinatorialOptimization1998,smithNeuralNetworksCombinatorial}. However, most problems of this class remain unsolvable in their full form due to the inefficiency of von Neumann computing architectures on solving them. On the other hand, the Ising machine hardware naturally solves combinatorial optimization problems by undergoing a dynamic energy minimization process, thereby achieving greater efficiency than existing digital computing schemes.

Ising machines have been realized based on many physical paradigms, including quantum annealing \cite{cohenQuantumAnnealingFoundations2015}, optical parametric oscillators \cite{inagakiCoherentIsingMachine2016}, electronic LC \cite{wangOIMOscillatorBasedIsing2019,chouAnalogCoupledOscillator2019} and phase-transition nano-oscillators \cite{duttaUnderstandingContinuousTimeDynamics2020,duttaIsingHamiltonianSolver2020}, and most recently spin-based oscillators \cite{houshangSpinHallIsing2020,albertssonUltrafastIsingMachines2021}. As opposed to quantum and optical implementations, Ising machines based on electrically coupled oscillators can be operated on chip-scale and at room temperature, making this approach most attractive for scalable computing hardware. Experimental demonstrations of the Ising machine using coupled kHz-MHz electronic oscillators have proven the ability of the oscillator network to solve the Ising model on millisecond time scales \cite{wangOIMOscillatorBasedIsing2019,chouAnalogCoupledOscillator2019,duttaUnderstandingContinuousTimeDynamics2020,duttaIsingHamiltonianSolver2020}. However, the Ising machine dynamics can be sped up by orders of magnitude using oscillators operating in the GHz frequency range, such as spin torque and spin Hall nano-oscillators (SHNOs). The sub-microsecond dynamics of the spin torque oscillator Ising machine were demonstrated in \cite{albertssonUltrafastIsingMachines2021} using a general numerical simulation framework that could be applied to any inter-oscillator coupling mechanism, such as electrical or magnetic. However, the phenomenological coupling strengths used in the model were not calculated from first principles using physical parameters, making it difficult to determine how the oscillator and system design will impact the coupled network performance. A previous experimental work further showed Ising-like collective dynamics in a small magnetically coupled network of oscillators \cite{houshangSpinHallIsing2020}. While the experimental realization of such a system is a significant milestone, an electrically coupled network will have significant advantages over the magnetic coupling scheme shown therein, as the latter only connects nearest-neighbor oscillators and the coupling strengths are not easily programmable post-device fabrication.

In this paper, we present an analytical and numerical study of the electrically coupled SHNO-based Ising machine for the advantages of reaching high connectivities among different nodes as well as tunable connection weights. We first introduce an analytical model that can describe the injection locking dynamics of a SHNO coupled to an external signal at the oscillator's fundamental frequency or harmonics. Different from previous approaches for studying the coupling of SHNOs, we develop a new approach -- using an impulse sensitivity function to describe the interactions of oscillators not only in small but also large angle oscillation regimes. Based on our analytical model, we develop a Verilog-A macromodel that can emulate the SHNO's injection locking behavior in a standard electrical circuit, which is validated by full micromagnetic simulations. We integrate our oscillator device model into circuit simulations with off-the-shelf electronic components to show how the Ising machine solution time and accuracy are expected to scale with the coupled array size and in the presence of phase noise. Our results provide analytical and quantitative tools to understand the performance of the electrically coupled SHNO network, enabling the future experimental realization of a fast, energy-efficient, and scalable spintronic Ising machine.

\section{\label{sec:Oscillator Ising Machine Theory}Oscillator Ising Machine}

\begin{figure}
 \includegraphics[width=7cm]{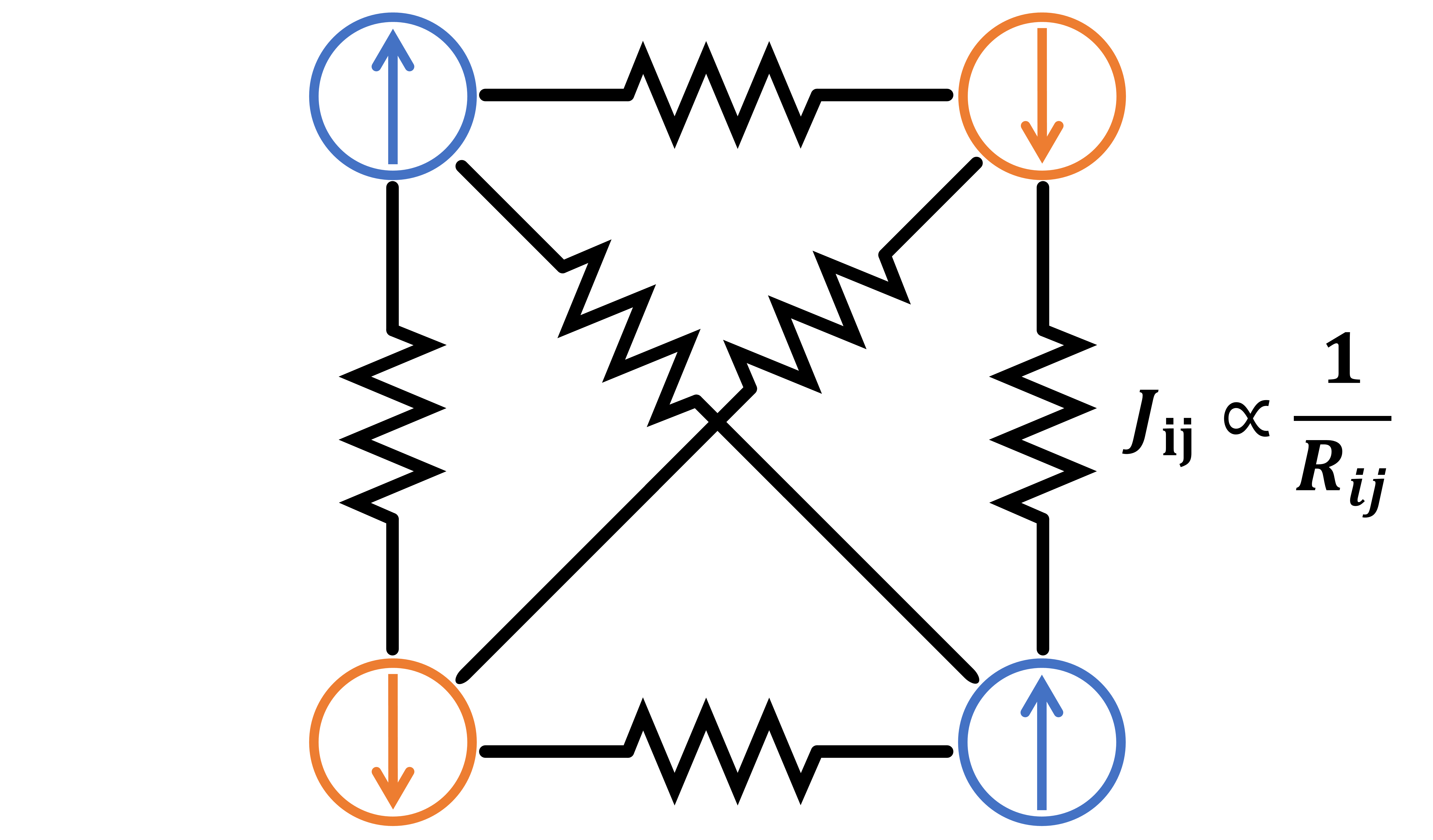}
 \caption{\label{fig:max_cut_example}Schematic of all-to-all electrically coupled network of 4 oscillators that can be mapped to the Ising model. Nodes represent oscillators that encode spin values $s_i=+1$ (``spin-up") or $s_i=-1$ (``spin-down") in their phases. For resistive coupling, the Ising model coefficients are proportional to the conductances linking each pair of oscillators \cite{wangOIMOscillatorBasedIsing2019}.}
\end{figure}

The mathematical Ising model describing domain formation in ferromagnets provides the theoretical basis for the Ising machine operation \cite{koesterBeitraegeZurTheorie2013}. The Ising Hamiltonian, simplified by neglecting the Zeeman term, is given below \cite{isingBeitragZurTheorie1925}
\begin{equation}
    H=-\sum_{i,j,i<j}^NJ_{ij}s_i s_j
    \label{eq:ising_model}
\end{equation}
where $s_i$ are discrete variables representing spins in a ferromagnetic material lattice and $J_{ij}$ are exchange coupling coefficients describing the interactions between spins. Under a given $\{J_{ij}\}$, spins take values $s_i=\pm1$ in order to minimize the system's total energy $H$. Many combinatorial optimization problems, including all 21 on Karp's well-known list of NP-hard problems, can be represented by an equivalent Ising Hamiltonian with spins encoding variables to optimize on, and the minimum-energy spin configuration representing the problem's optimal solution \cite{karpComputationalComplexityCombinatorial1975,lucasIsingFormulationsMany2014}.

A coupled network of nonlinear oscillators as shown in Fig. \ref{fig:max_cut_example} can be used to solve the Ising model in its collective phase dynamics. When a radio-frequency (rf) signal is applied to a nonlinear oscillator, the oscillator's frequency and/or phase can match that of the external signal by a process known as injection locking. Injection locking is described by the well-known Adler's equation, which can be generalized to describe a coupled network of $N$ oscillators \cite{adlerStudyLockingPhenomena1946,kuramotoCooperativeDynamicsOscillator1984}
\begin{equation}
    \frac{d\psi_i}{dt}=\omega_i-\sum_{j,j\neq i}^N K_{ij}\sin(\psi_i-\psi_j)
\end{equation}
where $\psi_i$ and $\omega_i$ are the phase and frequency of oscillator $i$ and $K_{ij}$ is the coupling strength between oscillators $i$ and $j$. By choosing a rotating frame $\phi_i=\psi_i-\omega^* t$ with respect to the mean oscillator frequency $\omega^*$, one further obtains
\begin{equation}
    \frac{d\phi_i}{dt}=\omega_i-\omega^*-\sum_{j,j\neq i}^N K_{ij}\sin(\phi_i-\phi_j)
    \label{eq:gen_adler}
\end{equation}

For coupled oscillators with close enough frequencies, each oscillator's phase will synchronize to its injected signal. This collective process is analogous to the minimization of the oscillator array's equivalent energy $H$, called the interaction potential, such that $\frac{d\phi_i}{dt}=-\frac{\partial H}{\partial\phi_i}$ \cite{shinomotoPhaseTransitionsActive1986,kuramotoCooperativeDynamicsOscillator1984}. The interaction potential for an oscillator array described by Eq. (\ref{eq:gen_adler}), neglecting the driving frequency difference, is given explicitly by
\begin{equation}
    H=-\sum_{i,j,i\neq j}^N K_{ij}\cos(\phi_i-\phi_j)
    \label{eq:interaction_potential}
\end{equation}
This takes a similar form to the Ising Hamiltonian in Eq. (\ref{eq:ising_model}) with coupling coefficients $K_{ij}=J_{ij}/2$ and oscillator phases representing spins $\cos(\phi_i-\phi_j)\mapsto s_i s_j=\pm1$. However, the oscillator phases are continuous variables whereas the Ising model spins are discrete. To resolve this difference, an effective uniaxial anistropy is introduced to the interaction Hamiltonian \cite{kuramotoCooperativeDynamicsOscillator1984}
\begin{equation}
    H=-\sum_{i,j,i\neq j}^N K_{ij}\cos(\phi_i-\phi_j)-K_s\sum_i^N \cos(2\phi_i)
    \label{eq:interaction_potential_shil}
\end{equation}
where the second term with coupling strength $K_s$ introduces energy minima at $\phi_i=\{0,\pi\}$. The oscillator phases are encouraged to binarize to these values representing discrete spins $s_i=\pm1$ in the Ising model. Correspondingly, Adler's equation becomes \cite{wangOIMOscillatorBasedIsing2019}
\begin{equation}
    \frac{d\phi_i}{dt}=\omega_i-\omega^*-\sum_{j,j\neq i}^N K_{ij}\sin(\phi_i-\phi_j)-2K_s \sin(2\phi_i)
    \label{eq:gen_adler_shil}
\end{equation}
where the binarizing term is physically realized by injecting a global second harmonic signal to the coupled oscillator network.

\section{\label{sec:Injection Locking Model of SHNO}Injection Locking Model}

\begin{figure*}
 \includegraphics[width=12.9cm]{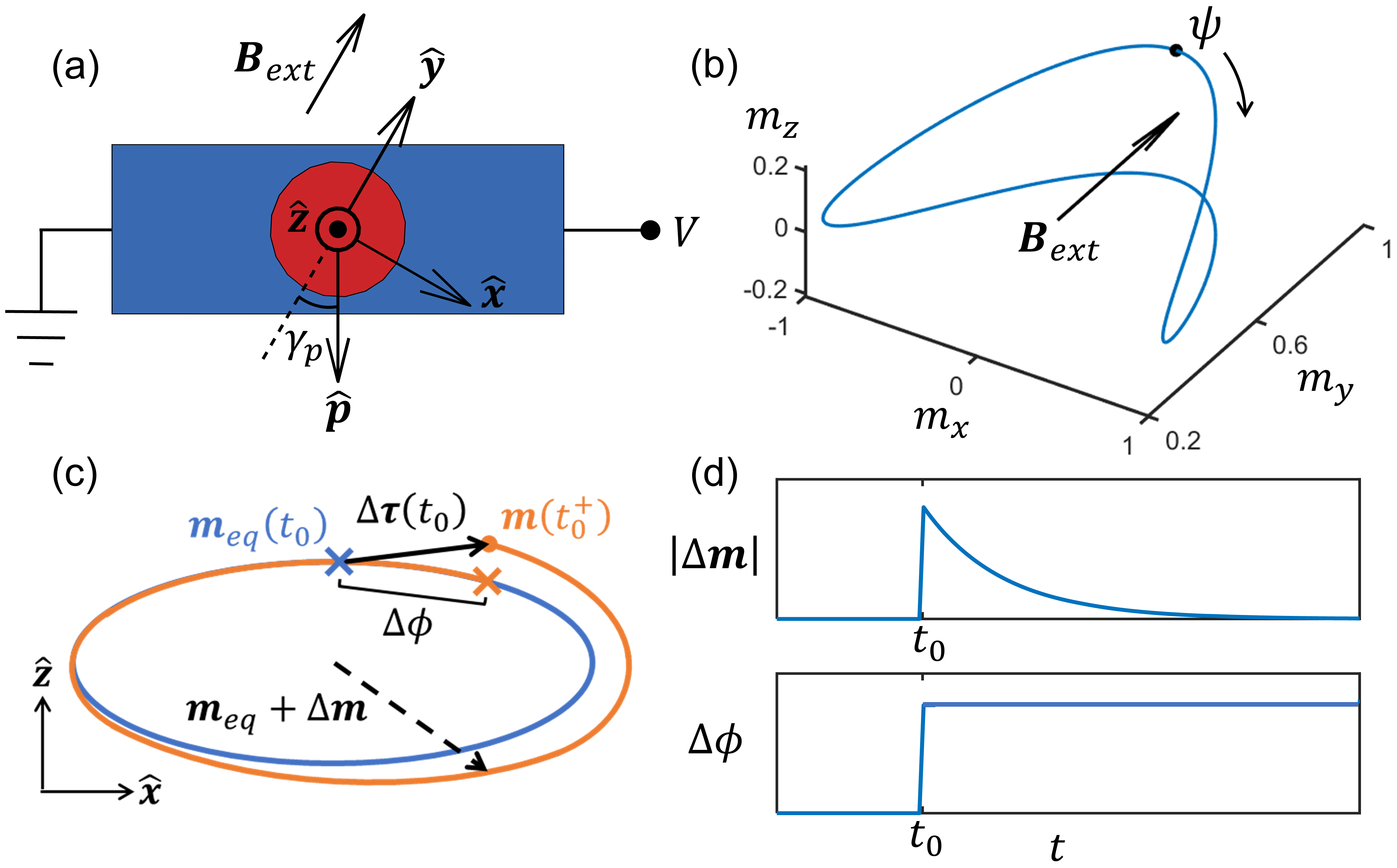}
 \caption{\label{fig:mtj_axis_def}  (a) Top view of 3-terminal SHNO. Blue is heavy metal and red is the magnetic tunnel junction (MTJ) stack (from top to bottom, consisting of a static magnetic fixed layer, insulating spacer layer, and precessing magnetic free layer adjacent to heavy metal). (b) Clamshell-shaped equilibrium precession $\mathbf{m}_{eq}$ of the in-plane magnetized SHNO calculated in micromagnetic simulation using parameters in Appendix \ref{app:Single Oscillator}. (c) Schematic of the SHNO's magnetization precession orbit subject to an impulse perturbation at time $t_0$ (blue cross). Blue (orange) orbits represent equilibrium (perturbed) precession. Orange cross shows phase deviation of oscillator one period after perturbation. (d) Amplitude and phase response of perturbed oscillator in (c).}
\end{figure*}

Spin Hall nano-oscillators, or more generally spin torque nano-oscillators, are a type of nonlinear oscillator with a nanometer footprint and GHz oscillation frequency originating from the direct-current-induced spontaneous magnetic precession (Figs. \ref{fig:mtj_axis_def}(a) and \ref{fig:mtj_axis_def}(b)) \cite{kiselevMicrowaveOscillationsNanomagnet2003,liuMagneticOscillationsDriven2012}. Generally, one relies on numerical solutions to describe the oscillation dynamics of the magnetic moments as well as their interactions with external signals. Previously, in the seminal work of \cite{slavinNonlinearSelfphaselockingEffect2005,slavinNonlinearAutoOscillatorTheory2009}, Slavin \textit{et al.} derived a compact analytical model describing injection locking of spin torque oscillators. Initially, this model only treated small angle precession in the near-threshold regime before being extended in \cite{serpicoTheoryInjectionLocking2009} to large angle circular precession trajectories. However, it still remains a challenging task to model injection locking in oscillators with non-trivial precession trajectories such as the ``clamshell" orbit. Moreover, the lack of freedom on the trajectory shape makes it challenging to account for injection locking at higher harmonic frequencies, which relies on the deviation of the magnetic moment precession from a perfect circular or elliptical shape.

Here, in order to characterize the injection locking behavior for practical SHNOs with large oscillation angles, we employ an alternative approach using the impulse sensitivity function (ISF) of the SHNO. The ISF is a characteristic function of a nonlinear oscillator describing the time-varying phase and amplitude sensitivity to perturbation by an external signal \cite{hajimiriGeneralTheoryPhase1998}. With the ISF, we derive an analytical expression for the injection locking strength of the SHNO based on the oscillation orbit, which can be used directly as an input parameter in Adler's equation. Our model comprehensively describes the injection locking behavior of a SHNO with any orbital trajectory and at both the fundamental and harmonic frequencies.

Despite being a nonlinear oscillator, the phase response of the SHNO to external perturbations at a given operating point can be described by an impulse response function. For a general nonlinear oscillator, the phase response to an impulse perturbation applied at time $t_0$ is \cite{hajimiriGeneralTheoryPhase1998,demirFloquetTheoryNonlinear2000,pepeEfficientLineartimeVariant2012}
\begin{equation}
    h_\phi(t,t_0)=\omega_g\Gamma(t_0)u(t-t_0)
    \label{eq:phase_response}
\end{equation}
where $\omega_g$ is the unperturbed oscillator frequency, $u(t-t_0)$ is a unit step, and the coefficient $\Gamma(t_0)$ is the ISF which reflects the oscillator's phase sensitivity to external perturbations over each period. This phase impulse response for the SHNO is illustrated in Figs. \ref{fig:mtj_axis_def}(c) and \ref{fig:mtj_axis_def}(d). Under this linear treatment of the phase response, the ISF includes all the information about the oscillator's nonlinearity through its dependence on the equilibrium oscillation orbit. The net phase change under an arbitrary perturbing signal $P(t)$ can be written as \cite{hajimiriGeneralTheoryPhase1998}
\begin{equation}
    \Delta\phi(t) = \int_{-\infty}^\infty h_\phi(t,\tau)P(\tau)d\tau=\omega_g\int_{-\infty}^t \Gamma(\tau)P(\tau)d\tau
    \label{eq:phase_response_integral}
\end{equation}
For the nonlinear SHNO, we can verify that this linear phase response assumption remains valid as long as the perturbing signal is not much larger than the DC biasing force (see discussion in Section \ref{sec:Correlation of ISF with Adler's Equation}). Meanwhile, amplitude deviations from equilibrium decay rapidly in time due to the restoring force of the nonlinear oscillator to its equilibrium orbit (Figs. \ref{fig:mtj_axis_def}(c) and \ref{fig:mtj_axis_def}(d)), and are therefore neglected \cite{hajimiriGeneralTheoryPhase1998,demirFloquetTheoryNonlinear2000,pepeEfficientLineartimeVariant2012}.

Previously, the ISFs of other oscillators have been calculated using various approaches such as nonlinear perturbation theory \cite{demirFloquetTheoryNonlinear2000,maffezzoniSynchronizationAnalysisTwo2010}, direct measurement of the oscillator's phase response to impulse perturbations \cite{pepeEfficientLineartimeVariant2012,kudoNumericalSimulationTemporal2010}, or analytically calculating the phase shifts by projecting the perturbation along the equilibrium oscillation trajectory \cite{hajimiriGeneralTheoryPhase1998}. We use the third method because it allows us to construct an intuitive expression for the ISF based on the SHNO's magnetization dynamics.

As a first step in deriving the ISF, we describe the equilibrium precession of the in-plane magnetized SHNO. The Landau-Lifshitz-Gilbert-Slonczewski (LLGS) equation is used to describe the magnetization precession dynamics \cite{landauTHEORYDISPERSIONMAGNETIC,gilbertPhenomenologicalTheoryDamping2004,slonczewskiCurrentdrivenExcitationMagnetic1996}
\begin{equation}
    \begin{aligned}
    \frac{d\mathbf{m}}{dt}=-\frac{\gamma_e}{1+\alpha^2} \big[\mathbf{m}\times \mathbf{B}_{eff} + \alpha \big(\mathbf{m}\times(\mathbf{m}\times \mathbf{B}_{eff})\big)&\\
    +B_s\big(\mathbf{m}\times(\mathbf{m}\times \mathbf{\hat{p}})\big) - \alpha B_s(\mathbf{m}\times\mathbf{\hat{p}}&)\big]
    \end{aligned}
    \label{eq:llgs_full}
\end{equation}
where $\gamma_e$ is the electron gyromagnetic ratio, $\alpha$ is the Gilbert damping parameter, $\mathbf{m}$ is the normalized free layer magnetization, $\mathbf{B}_{eff}$ is the effective magnetic field, $B_s$ is the spin torque field, and $\mathbf{\hat{p}}$ is the injected spin polarization direction. The spin torque term proportional to $\alpha B_s$ has the same symmetry as the field-like torque and is usually neglected. Under stable oscillation, the torque from the DC current counterbalances the magnetic damping and maintains the equilibrium oscillation orbit. Therefore, we can separate the torques on the RHS of Eq. (\ref{eq:llgs_full}) into two parts,
\begin{equation}
    \begin{aligned}
    \frac{d\mathbf{m}}{dt}=-\gamma_e\big[&\mathbf{m}\times \mathbf{B}_{eff} + \alpha \big(\mathbf{m}\times(\mathbf{m}\times \mathbf{B}_{eff})\big)\\
    +&(B_{s0}+b_{srf})\big(\mathbf{m}\times(\mathbf{m}\times \mathbf{\hat{p}})\big)\big]\\
    =\boldsymbol{\tau}_{eq}+&\boldsymbol{\tau}_{rf}
    \end{aligned}
    \label{eq:llgs}
\end{equation}
where the DC and rf components of $B_s$ are separated based on the voltages applied to the heavy metal in Fig. \ref{fig:mtj_axis_def}(a)
\begin{equation}
    (B_{s0}+b_{srf})\mathbf{\hat{p}}=\frac{\sigma}{R_{HM}}\big(V_{DC}+V_{rf}\big)
    \begin{pmatrix}
        \sin(\gamma_p) \\
        -\cos(\gamma_p) \\
        0
    \end{pmatrix}
    \label{eq:Bst_1f}
\end{equation}
with $\sigma=\hbar\theta_{SH}/(2eM_s A t_{FL})$ where $\theta_{SH}$ is the spin Hall angle, $M_s$ is the free layer saturation magnetization, $A$ is the cross-sectional area of the heavy metal, and $t_{FL}$ is the free layer thickness. $R_{HM}$ is the total resistance of the heavy metal strip and $\gamma_p$ is the spin polarization angle shown in Fig. \ref{fig:mtj_axis_def}(a). A voltage bias applied across the MTJ for readout will introduce an additional DC spin transfer torque term into Eq. (\ref{eq:llgs}); however, this is estimated to be an order of magnitude smaller than the spin orbit torque excitation due to $V_{DC}$, and is therefore neglected.

In describing the equilibrium precession, we assume the rf spin torque field $b_{srf}$ is relatively weak compared to the DC component $B_{s0}$. One can then numerically solve the LLGS equation considering only $\boldsymbol{\tau}_{eq}$ in macrospin or micromagnetic simulations to find the equilibrium precession (see Appendix \ref{app:Single Oscillator} for our simulation details). A typical equilibrium orbit is shown in Fig. \ref{fig:mtj_axis_def}(b). To leading order, this can be approximated by
\begin{equation}
    \mathbf{m}_{eq}=
    \begin{pmatrix}
        m_x\sin(\psi) \\
        m_{y0}+m_y\cos(2\psi) \\
        m_z\cos(\psi)
    \end{pmatrix}
    \label{eq:m_eq}
\end{equation}
with the total oscillator phase $\psi=\omega_g t+\phi$. Notably, due to the clamshell-shaped precession, the magnetization $y$ component oscillates at twice the frequency of the $x$ and $z$ components. In following sections, we will see that this enables strong injection locking at the second harmonic.

\subsection{\label{sec:Impulse Sensitivity Function} Impulse Sensitivity Function}

We derive the ISF by analytically projecting the spin torque perturbation along the equilibrium precession trajectory to calculate the resulting phase shifts. This technique was originally proposed by Hajimiri \textit{et al.} for general nonlinear oscillators \cite{hajimiriGeneralTheoryPhase1998}. Here, we extend this method to the SHNO based on the magnetization dynamics described by the LLGS equation. Our general expression is used to predict the injection locking sensitivity to an applied time-varying voltage.

\begin{figure*}
 \includegraphics[width=17.2cm]{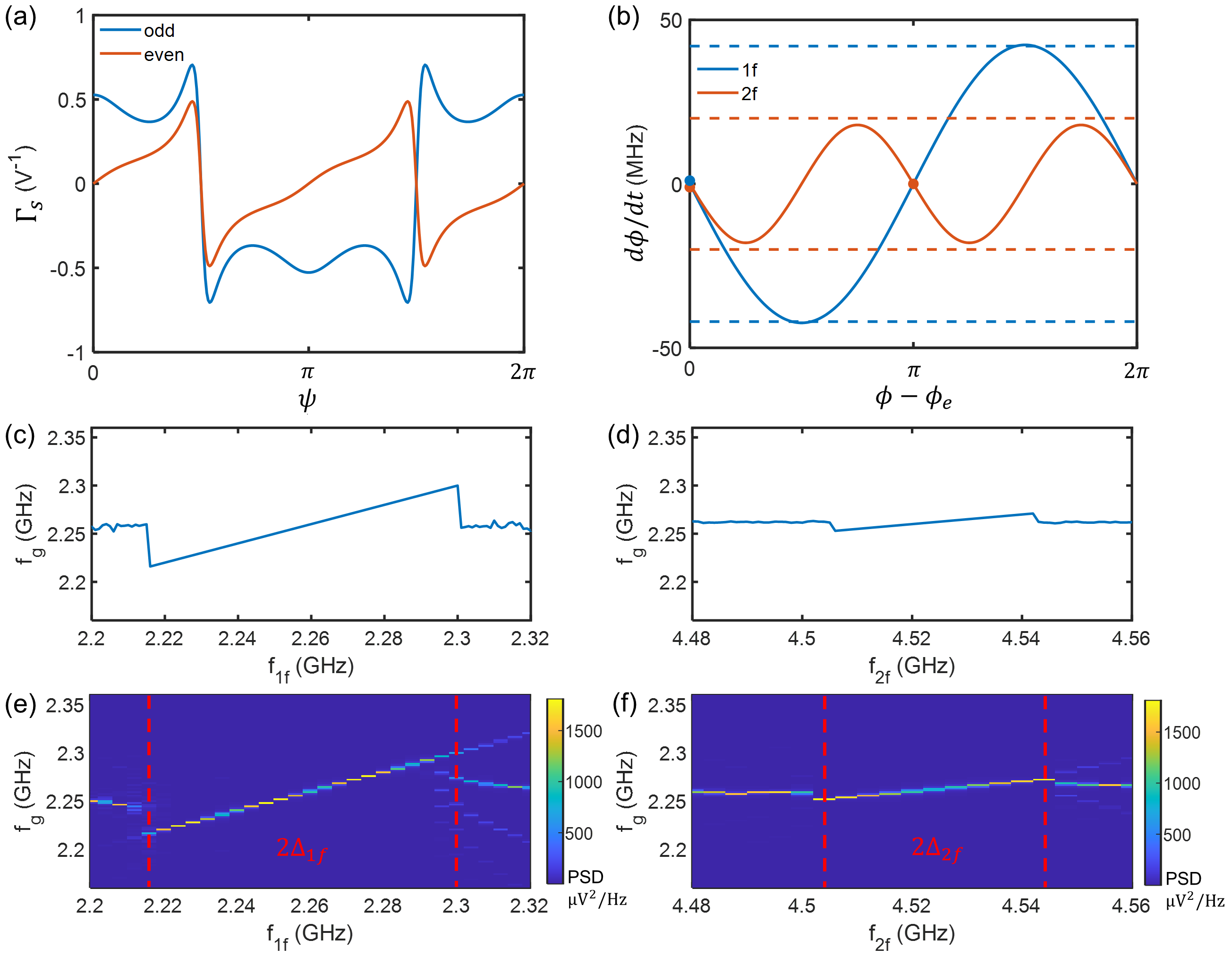}
 \caption{\label{fig:isf_adler_schematic}ISF and Adler's equation for device with parameters given in Appendix \ref{app:Single Oscillator}. (a) Odd and even series of the SHNO ISF (Eq. (\ref{eq:isf_rf_fourier})) based on a typical clamshell orbit. (b) Solid lines: Adler's equation (Eq. (\ref{eq:isf_adler})) with $V_{1f}=-V_{2f}=66$ mV. The $\mathrm{2f}$ component's amplitude is doubled to account for the second harmonic frequency. Dotted lines: Phase locking bandwidths $\Delta=|d\phi/dt|$ calculated in micromagnetic simulations (e,f). The injection locking strength predicted by Eq. (\ref{eq:isf_adler}) very closely matches the simulated phase locking bandwidth, showing that our ISF model can accurately predict injection locking at the first and second harmonics. We can also tell the stable phase locking angles (marked by dots) between the oscillator and external signal where $d\phi/dt=0$ and $d^2\phi/dt^2<0$. At the first harmonic, the oscillator is expected to lock in phase with the external signal, while the second harmonic signal gives bistable locking that binarizes the phases as described in Section \ref{sec:Oscillator Ising Machine Theory}. (c,d) Phase locking bandwidths calculated by numerically integrating Eq. (\ref{eq:isf_adler}) with $V_{rf}=66$ mV at (c) first harmonic $f_{1f}\approx f_g$ and (d) second harmonic $f_{2f}\approx 2f_g$. (e,f) Power spectral density of SHNO calculated in micromagnetic simulations under injection locking at (e) first harmonic and (f) second harmonic. The phase locking bandwidths are marked by vertical dotted lines, showing the boundaries of the region where the oscillator phase locks to the external signal.}
\end{figure*}

First, consider a SHNO with equilibrium magnetization precession given by $\mathbf{m}_{eq}(t)$ in Eq. (\ref{eq:m_eq}). Considering a perturbing impulse torque $\Delta\boldsymbol{\tau}(t)\delta(t-t_0)$, we expect a response of the oscillator magnetization $\mathbf{m}(t_0^+)=\mathbf{m}_{eq}(t_0)+\Delta\boldsymbol{\tau}(t_0)$ (Fig. \ref{fig:mtj_axis_def}(c)). Because we neglect amplitude deviations, we only consider the component of $\Delta\boldsymbol{\tau}$ along the equilibrium precession trajectory. The displacement $l$ along the precession trajectory is then given by $\Delta\boldsymbol{\tau}$ dotted with a unit vector along the trajectory direction
\begin{equation}
    l=\Delta\boldsymbol{\tau}(t)\cdot\frac{\dot{\mathbf{m}}_{eq}(t)}{|\dot{\mathbf{m}}_{eq}(t)|}
\end{equation}
We then divide by the instantaneous precession speed $|\dot{\mathbf{m}}_{eq}|$ to find the equivalent phase shift in units of time
\begin{equation}
    \Delta \phi_t=\Delta\boldsymbol{\tau}(t)\cdot\frac{\dot{\mathbf{m}}_{eq}(t)}{|\dot{\mathbf{m}}_{eq}(t)|^2}
\end{equation}
Normalizing the phase shift with the magnitude of the perturbation, we obtain the ISF for a SHNO perturbed by a voltage $V$
\begin{equation}
    \Gamma_s(t)=\frac{\Delta\boldsymbol{\tau}(t)}{V}\cdot\frac{\dot{\mathbf{m}}_{eq}(t)}{|\dot{\mathbf{m}}_{eq}(t)|^2}
    \label{eq:shno_isf_rf}
\end{equation}

For the stable solution in Eq. (\ref{eq:m_eq}), one can calculate the ISF with Eq. (\ref{eq:shno_isf_rf}) and express it in the form of a Fourier series
\begin{equation}
\begin{aligned}
    \Gamma_{s}(t)=\frac{\omega_g\gamma_e\sigma}{4 R_{HM}|\dot{\mathbf{m}}_{eq}|^2}&\bigg[\sum_{n=\textrm{odd}} p_x a_n\frac{1}{n^2}\cos\big(n(\omega_g t+\phi)\big)\\
    &+\sum_{n=\textrm{even}} p_y b_n\frac{1}{n^2}\sin\big(n(\omega_g t+\phi)\big)\bigg]
    \label{eq:isf_rf_fourier}
\end{aligned}
\end{equation}
with non-zero coefficients
\begin{equation*}
    \begin{aligned}
    a_1=&2m_x\big(m_y^2+4 m_y m_{y0}+2(m_{y0}^2+m_z^2)\big) \\
    a_3=&3m_x m_y^2 \\
    a_5=&-m_x m_y^2 \\
    b_2=&-2\big(m_x^2(2m_y+m_{y0})+m_z^2(2m_y-m_{y0})\big) \\
    b_4=&m_y(m_x^2-m_z^2)
\end{aligned}
\end{equation*}
where the instantaneous precession speed is
\begin{equation}
\begin{aligned}
    |\dot{\mathbf{m}}_{eq}|^2=\frac{\omega_g^2}{2}&\big[(m_x^2+4m_y^2+m_z^2)\\
    &+(m_x^2-m_z^2)\cos\big(2(\omega_g t+\phi)\big)\\
    &-4m_y^2\cos\big(4(\omega_g t+\phi)\big)\big]
    \label{eq:m_eq_2nd_order}
\end{aligned}
\end{equation}
Fig. \ref{fig:isf_adler_schematic}(a) shows the ISF calculated for a SHNO with a typical clamshell oscillation orbit. Here, we have plotted separately the odd- and even-harmonic series of Eq. (\ref{eq:isf_rf_fourier}) proportional to spin polarization components that are asymmetric ($p_x$) and symmetric ($p_y$) with respect to the precession trajectory \cite{urazhdinFractionalSynchronizationSpinTorque2010}.

\subsection{\label{sec:Correlation of ISF with Adler's Equation}Adler's Equation}

The ISF contains all necessary information for describing injection locking of the SHNO, allowing us to relate the Adler's equation coefficients (Eq. (\ref{eq:gen_adler_shil})) with the oscillator's physical properties. We define $\phi(t)$ as the phase in the rotating frame as in Eq. (\ref{eq:gen_adler}), and from Eq. (\ref{eq:phase_response_integral}), we have:
\begin{equation}
    \phi(t)=(\omega_g-\omega_e)t+\omega_g\int_{-\infty}^t\Gamma_s(\tau)P(\tau)d\tau
    \label{eq:delta_phi}
\end{equation}
where $\omega_e$ is the external signal frequency. Calculating the derivative with respect to $t$ on both sides of Eq. (\ref{eq:delta_phi}), we obtain
\begin{equation}
    \frac{d\phi(t)}{dt}=\omega_g-\omega_e+\omega_g\Gamma_s(t)P(t)
\end{equation}
Because the phase evolution $d\phi/dt$ that contributes to the long term dynamics is much slower than the oscillatory dynamics, we can approximate the right side by the average over one period \cite{kuramotoCooperativeDynamicsOscillator1984}
\begin{equation}
    \frac{d\phi(t)}{dt}=\omega_g-\omega_e+\frac{\omega_g}{T}\int_0^T \Gamma_s(t)P(t)dt
    \label{eq:adler_xcorr}
\end{equation}

Consider an injection locking signal that contains both first and second harmonic components
\begin{equation}
    P(t)=V_{1f}\sin\big(\omega_e t+\phi_{e1}\big)+V_{2f}\cos\big(2(\omega_e t+\phi_{e2})\big)
    \label{eq:p_inj_lock}
\end{equation}
Plugging Eqs. (\ref{eq:isf_rf_fourier}) and (\ref{eq:p_inj_lock}) into Eq. (\ref{eq:adler_xcorr}), we see that the only relevant terms of $\Gamma_s$ that contribute to the long term phase variations are
\begin{figure}
 \includegraphics[width=8.6cm]{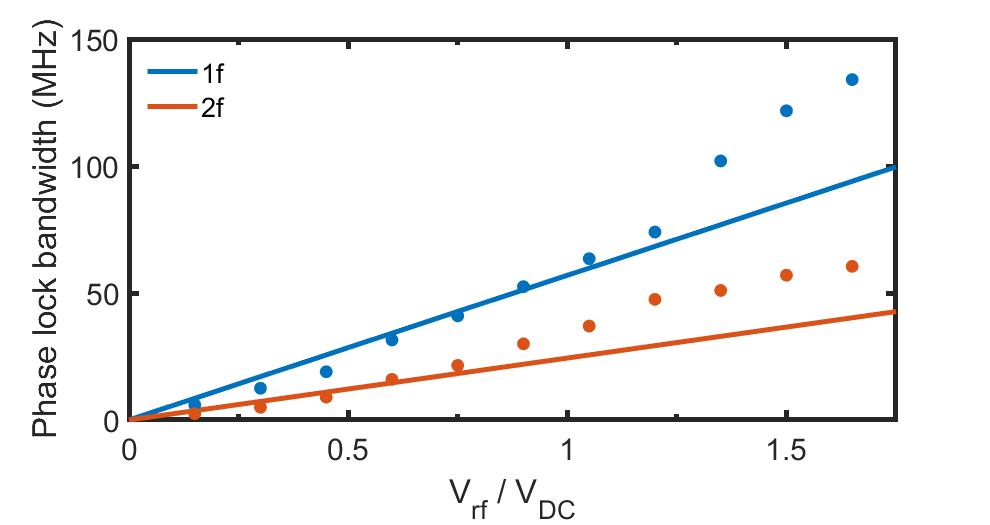}
 \caption{\label{fig:isf_linear_approx}Phase locking bandwidth as a function of injection locking signal strength at first and second harmonics. Lines represent bandwidths calculated from ISF (coefficients of Eq. (\ref{eq:isf_adler})), while scatters are the corresponding bandwidths obtained from micromagnetic simulations based on device parameters in Appendix \ref{app:Single Oscillator}.}
\end{figure}
\begin{equation}
\begin{aligned}
    \Gamma_s(t)=&\frac{\gamma_e\sigma p_x}{2\omega_g R_{HM}}a_1'\cos(\omega_e t+\phi)\\
    &+\frac{\gamma_e\sigma p_y}{4\omega_g R_{HM}}b_2'\sin(2(\omega_e t+\phi))\\
    =&\Gamma_{1f}(t)+\Gamma_{2f}(t)
    \label{eq:isf_approx_1f_2f}
\end{aligned}
\end{equation}
where $a_1'$ and $b_2'$ are calculated by approximating Eqs. (\ref{eq:isf_rf_fourier}) and (\ref{eq:m_eq_2nd_order}) to second order and finding the Fourier series coefficients
\begin{equation*}
    \begin{aligned}
    a_1'=&\frac{x_1}{y}\bigg(1-\sqrt{\bigg|\frac{y-1}{y+1}\bigg|}\bigg) \\
    b_2'=&\frac{x_2}{y^2}\bigg(1-\sqrt{\big|y^2-1\big|}\bigg) \\
    x_1=&\frac{a_1}{m_x^2+4m_y^2+m_z^2} \\
    x_2=&\frac{b_2}{m_x^2+4m_y^2+m_z^2} \\
    y=&\frac{m_x^2-m_z^2}{m_x^2+4m_y^2+m_z^2}
\end{aligned}
\end{equation*}
Note that the higher order terms of $\Gamma_s$ in Eq. (\ref{eq:isf_rf_fourier}) lead to fast phase oscillations which do not accumulate beyond one period and are therefore neglected. Finally, evaluating Eq. (\ref{eq:adler_xcorr}) gives us Adler's equation for the SHNO
\begin{equation}
    \begin{aligned}
    \frac{d\phi(t)}{dt}=\omega_g-\omega_e-&\frac{\gamma_e\sigma p_x }{4 R_{HM}}a_1'V_{1f}\sin\big(\phi-\phi_{e1}\big)\\
    +&\frac{\gamma_e\sigma p_y }{4 R_{HM}}b_2'V_{2f}\sin\big(2(\phi-\phi_{e2})\big)
    \label{eq:isf_adler}
\end{aligned}
\end{equation}
Comparing to Eq. (\ref{eq:gen_adler_shil}), we have extracted the coupling coefficients of Adler's equation in terms of the material and device parameters that possess real physical meanings. The first and second harmonic terms of Eq. (\ref{eq:isf_adler}) are plotted separately in Fig. \ref{fig:isf_adler_schematic}(b). While this equation describes the injection locking of a single oscillator, it can be trivially extended to a coupled network of oscillators as in Eq. (\ref{eq:gen_adler_shil}) (see Section \ref{sec:Electrically Coupled Oscillator Array}).

To demonstrate that our model can accurately predict first and second harmonic injection locking, we numerically evaluate Eq. (\ref{eq:isf_adler}) to find the injection locked oscillator's frequency as a function of $f_e=\omega_e/2\pi$. We apply an rf perturbing signal $P(t)=V_{rf}\sin(\omega_e t)$ to the free-running oscillator and sweep the frequency $\omega_e$ about $\omega_g$ and $2\omega_g$ for first and second harmonic injection locking, respectively. We obtain the results in Figs. \ref{fig:isf_adler_schematic}(c) and \ref{fig:isf_adler_schematic}(d). We then carry out micromagnetic simulations with the same device parameters and rf voltages. In Figs. \ref{fig:isf_adler_schematic}(e) and \ref{fig:isf_adler_schematic}(f), we show the power spectral density of the oscillator's microwave output voltage as a function of $f_e$, from which one sees that Eq. (\ref{eq:isf_adler}) and the full micromagnetic simulations lead to the same calculated coupling bandwidths. Furthermore, we evaluate the coupling bandwidths as a function of the injection locking signal amplitude in Fig. \ref{fig:isf_linear_approx}, comparing our analytical model and micromagnetic simulations. For coupling signal amplitudes up to $V_{rf}\approx V_{DC}$, the bandwidths calculated in micromagnetic simulations follow the linear trend predicted by Eq. (\ref{eq:isf_adler}), demonstrating that our ISF-based model holds across a wide range of coupling signals.

\section{\label{sec:Collective Phase Dynamics of Electrically Coupled Oscillator Network}Circuit-Level Simulation}

\begin{figure}
 \centering
 \includegraphics[width=8.6cm]{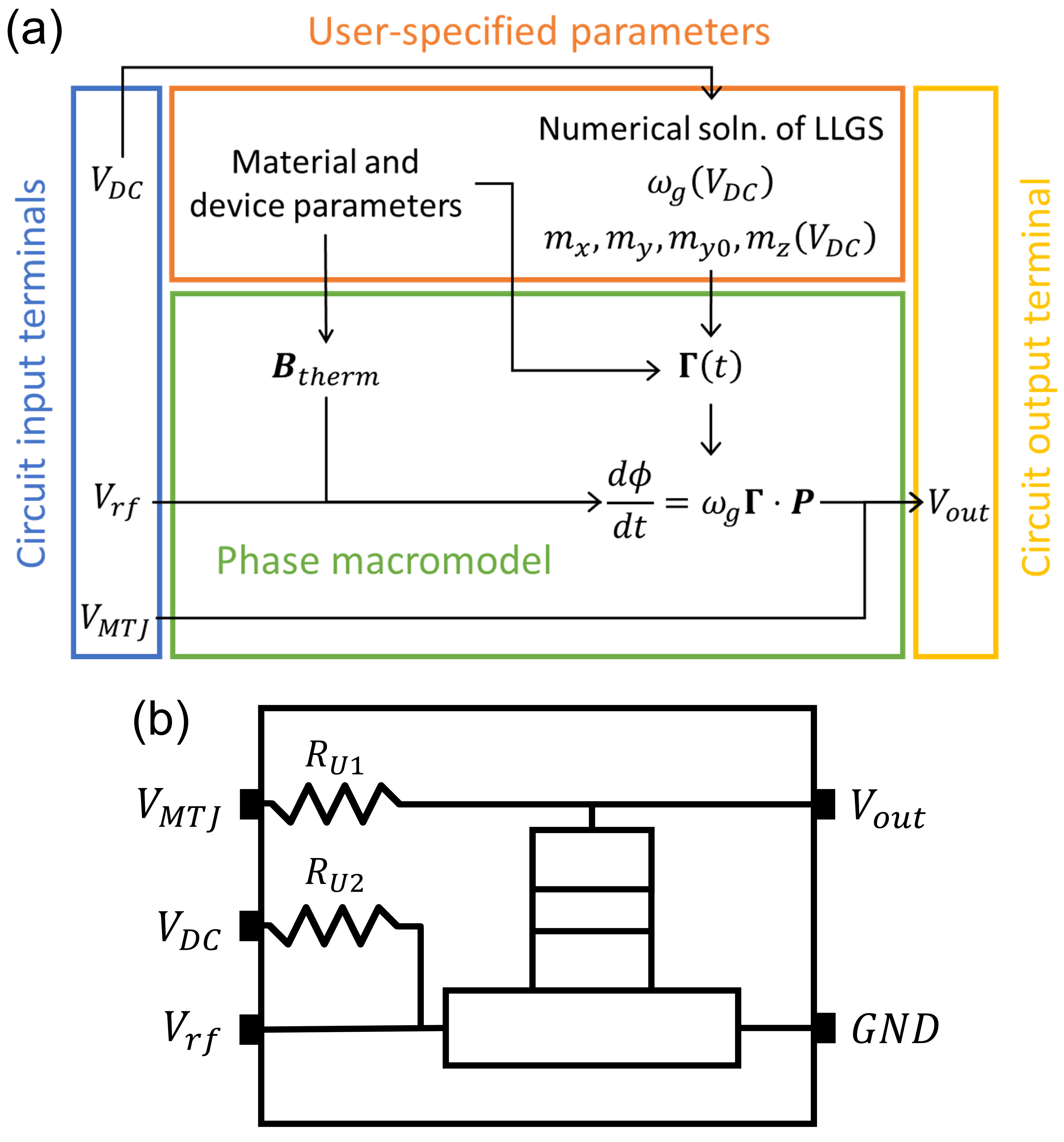}
 \caption{\label{fig:macromodel_package}(a) Block diagram of SHNO macromodel implemented in Verilog-A, including electrical behavior and nonlinear phase macromodel to calculate injection locking and phase noise. (b) Circuit-level symbol of SHNO showing electrical terminals and pull-up resistors $R_{U1},R_{U2}$.}
\end{figure}

\begin{table}
\caption{\label{tab:va_parameters}
User-specified material and device parameters in the Verilog-A SHNO macromodel, including frequency and equilibrium precession components at the given operating point $V_{DC}$. The values given are those used in our simulations.}
\begin{ruledtabular}
\begin{tabular}{llll}
 Parameter & Symbol & Value & Unit \\
\hline
Parallel MTJ resistance & $R_P$ & 1200 & $\Omega$ \\
Anti-parallel MTJ resistance & $R_{AP}$ & 2400 & $\Omega$ \\
Heavy metal resistance & $R_{HM}$ & 200 & $\Omega$ \\
Heavy metal width & $w_{HM}$ & 200 & nm \\
Heavy metal thickness & $t_{HM}$ & 5 & nm \\
Spin polarization angle & $\gamma_p$ & 30 & $^{\circ}$ \\
Spin Hall angle & $\theta_{SH}$ & 0.07 & - \\
Gilbert damping & $\alpha$ & 0.02 & - \\
Saturation magnetization & $M_s$ & 800 & kA/m \\
MTJ width & $w_{MTJ}$ & 120 & nm \\
MTJ length & $l_{MTJ}$ & 200 & nm \\
Free layer thickness & $t_{FL}$ & 1.5 & nm \\
Thermal field timestep & $\Delta t$ & 15 & ps \\
MTJ voltage & $V_{MTJ}$ & 0.4 & V \\
MTJ pull-up resistor & $R_{U1}$ & 1.9 & k$\Omega$ \\
DC voltage & $V_{DC}$ & 0.968 & V \\
Heavy metal pull-up resistor & $R_{U2}$ & 2.0 & k$\Omega$ \\
Free-running frequency & $\omega_g(V_{DC})$ & 2.26 & GHz \\
Components of $\mathbf{m}_{eq}(V_{DC})$ & $m_x$ & 0.99 & - \\
& $m_y$ & 0.34 & - \\
& $m_{y0}$ & 0.6 & - \\
& $m_z$ & 0.21 & - \\
\end{tabular}
\end{ruledtabular}
\end{table}

To reliably predict the performance of the SHNO Ising machine, we need to simulate large arrays up to hundreds or thousands of coupled oscillators. To do so, we develop a computationally lightweight SHNO macromodel based on our analytical ISF. The oscillator macromodel is integrated with off-the-shelf electronic components in circuit simulations to realize a tunable electrical coupling scheme. We show that our macromodel can accurately replicate the collective phase dynamics of a SHNO array calculated in micromagnetic simulations but in a very small percentage of the computation time.

\subsection{\label{sec:SHNO Device Model in Verilog-A}Verilog-A Oscillator Macromodel}

We introduce an ISF-based macromodel of the SHNO in Verilog-A that emulates the oscillator's electrical behavior, nonlinear phase dynamics, and thermal phase noise characteristic \cite{popescuSimulationCoupledSpin2018,guImplementingNonlinearOscillator2005,xiaoluelaiCapturingOscillatorInjection2004,xiaoluelaiFastAccurateSimulation2005,maffezzoniPhaseNoiseReductionOscillators2011}. Previous approaches to model the spin torque oscillator at the circuit level have involved numerically simulating the LLGS equation using an equivalent circuit representation \cite{csabaModelingCoupledSpin2012,panagopoulosPhysicsBasedSPICECompatibleCompact2013,amentSolvingStochasticLandauLifshitzGilbertSlonczewski2017}, which is computationally expensive, or solving analytical equations that accurately model the electrical behavior, including output power and linewidth, but do not include nonlinear injection locking \cite{ahnAnalyticModelSpinTorque2013,chenComprehensiveMacrospinBasedMagnetic2015,yogendraCoupledSpinTorqueNanoOscillatorBased2017,kazemiCompactModelSpin2016,albertssonCompactMacrospinBasedModel2019}. On the other hand, our analytical ISF-based approach comprehensively models the nonlinear behavior of the oscillator without requiring heavy computation. The macromodel code described below is open-source and available online \footnote{Brooke McGoldrick. Verilog-A SHNO Macromodel. \url{https://github.com/bcmcgold/va-shno}.}.

A functional block diagram of the SHNO macromodel is shown in Fig. \ref{fig:macromodel_package}(a). The user-input material and device parameters are specified in Table \ref{tab:va_parameters} with values given for the SHNO studied here. The oscillator's free-running angular frequency $\omega_g$ and components of $\mathbf{m}_{eq}$ (to leading order, as in Eq. (\ref{eq:m_eq})) in the top block of Fig. \ref{fig:macromodel_package}(a) can be input as a function of $V_{DC}$ to model the oscillator across a range of operating points. These values can be extracted from a one-time micromagnetic or macrospin simulation of the SHNO device and imported to Verilog-A as look-up tables.

In the initial circuit simulation timestep, the oscillator macromodel undergoes a DC operating point analysis. First, $V_{DC}$ is sampled to determine $\omega_g$ and $\mathbf{m}_{eq}$ at the given operating point. Then, the ISF is calculated from Eq. (\ref{eq:isf_rf_fourier}). While we have only provided the ISF for in-plane precession, a similar derivation as in Section \ref{sec:Impulse Sensitivity Function} can be carried out for any arbitrary precession orbit and coded into Verilog-A.

Besides injection locking by deterministic electrical signals, the ISF approach can be used to model phase noise due to stochastic thermal fluctuations \cite{maffezzoniPhaseNoiseReductionOscillators2011}. Thermal fluctuations can be treated as an effective field in the macrospin approximation \cite{brownThermalFluctuationsSingleDomain1963,xiaoMacrospinModelsSpin2005}
\begin{equation}
    \mathbf{B}_{therm}=\boldsymbol{\eta}(t)\sqrt{\frac{2\alpha k_B T}{\gamma_e V_{FL} M_s \Delta t}}
    \label{eq:b_therm}
\end{equation}
where $\boldsymbol{\eta}(t)$ is a 3D Cartesian vector with components randomly chosen from a normal distribution every $\Delta t$, $T$ is the temperature, and $V_{FL}$ is the volume of the oscillator's free layer. Because the ISF is a characteristic of the oscillator and independent of the perturbation waveform, the effective thermal field ISF is derived by a similar process as for the spin torque ISF. The ISFs corresponding to perturbation by effective magnetic fields along $x$, $y$, and $z$ directions are calculated in Appendix \ref{sec:Magnetic Field Impulse Sensitivity Function} and integrated into the oscillator macromodel in analytical form.

As we showed in Section \ref{sec:Correlation of ISF with Adler's Equation}, the dynamic phase behavior of the SHNO is well captured by Adler's equation with our analytically derived coupling coefficients. Because Adler's equation is a simple first order differential equation, it can be solved very efficiently in SPICE-based circuit simulators in the form shown in Fig. \ref{fig:macromodel_package}(a), comprising the phase macromodel. The phase responses to spin torque and thermal field perturbations are calculated in parallel by multiplying each ISF with the corresponding perturbation and summing the total contribution to the phase at each timestep.

Finally, we package the analysis above into a compact SHNO device model as shown in Fig. \ref{fig:macromodel_package}(b). The device interacts with the external circuit only through its input terminals including MTJ bias voltage $V_{MTJ}$, DC driving voltage $V_{DC}$, and rf locking signal $V_{rf}$, as well as its output terminal $V_{out}$. The effective resistance across each pair of electrical terminals is modeled based on the heavy metal and MTJ resistances, including the rf oscillating magnetoresistance of the nanopillar. These resistances are used to construct the MTJ's electrical output signal
\begin{equation}
\begin{aligned}
    V_{out}=\frac{V_{MTJ}}{R_{MTJ}+R_{U1}}&\big(R_{MTJ}\\
    +\frac{R_P-R_{AP}}{2}&m_x\sin(\omega_g t+\phi(t))\big)
    \label{eq:vout}
\end{aligned}
\end{equation}
where $R_{MTJ}=(R_{AP}+R_P+R_{HM})/2$ is the equivalent DC resistance between the MTJ readout terminal and ground. Here we assume that the fixed layer of the MTJ is oriented along the $+x$ axis due to shape anisotropy.

\subsection{\label{sec:Electrically Coupled Oscillator Array}Coupled Oscillator Array}

\begin{figure}
 \includegraphics[width=8.6cm]{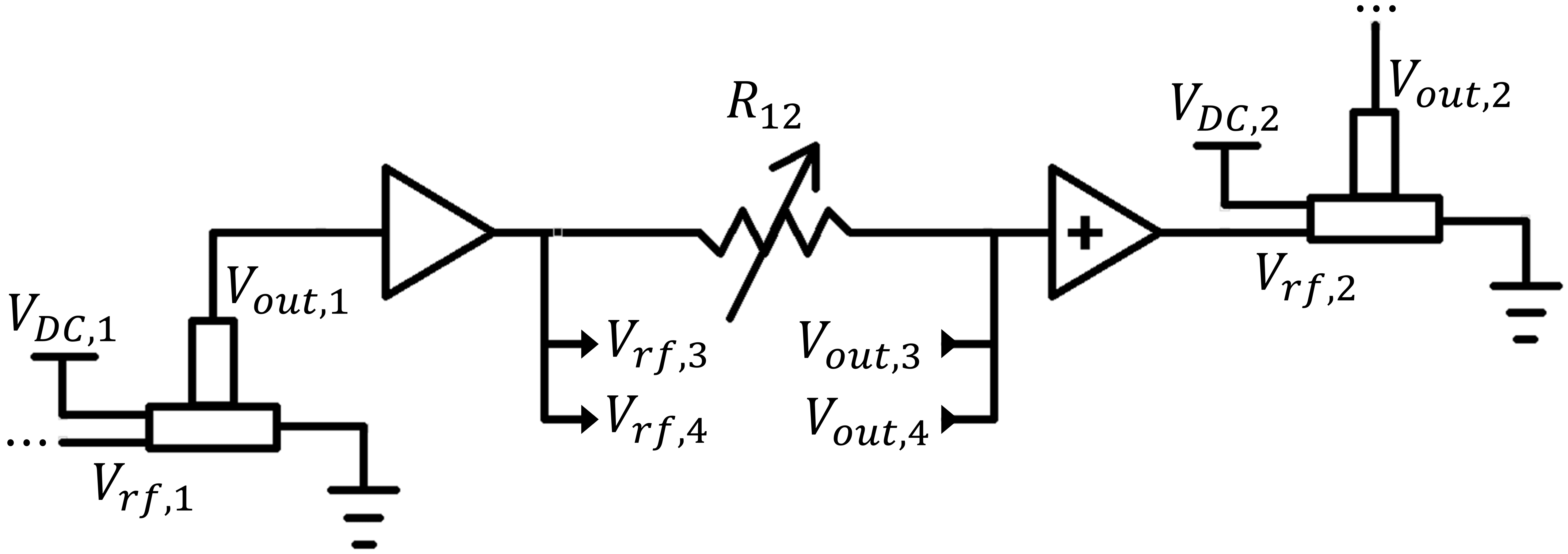}
 \caption{\label{fig:simp_circ_diagram}HSPICE coupling circuit schematic using SHNO macromodel and off-the-shelf amplifier models. One full coupling branch is shown for the all-to-all connected network in Fig. \ref{fig:max_cut_example} with oscillators arbitrarily numbered 1-4. Auxiliary gain-setting resistors and DC biasing elements are omitted for visual clarity.}
\end{figure}

\begin{figure}
 \centering
 \includegraphics[width=8.6cm]{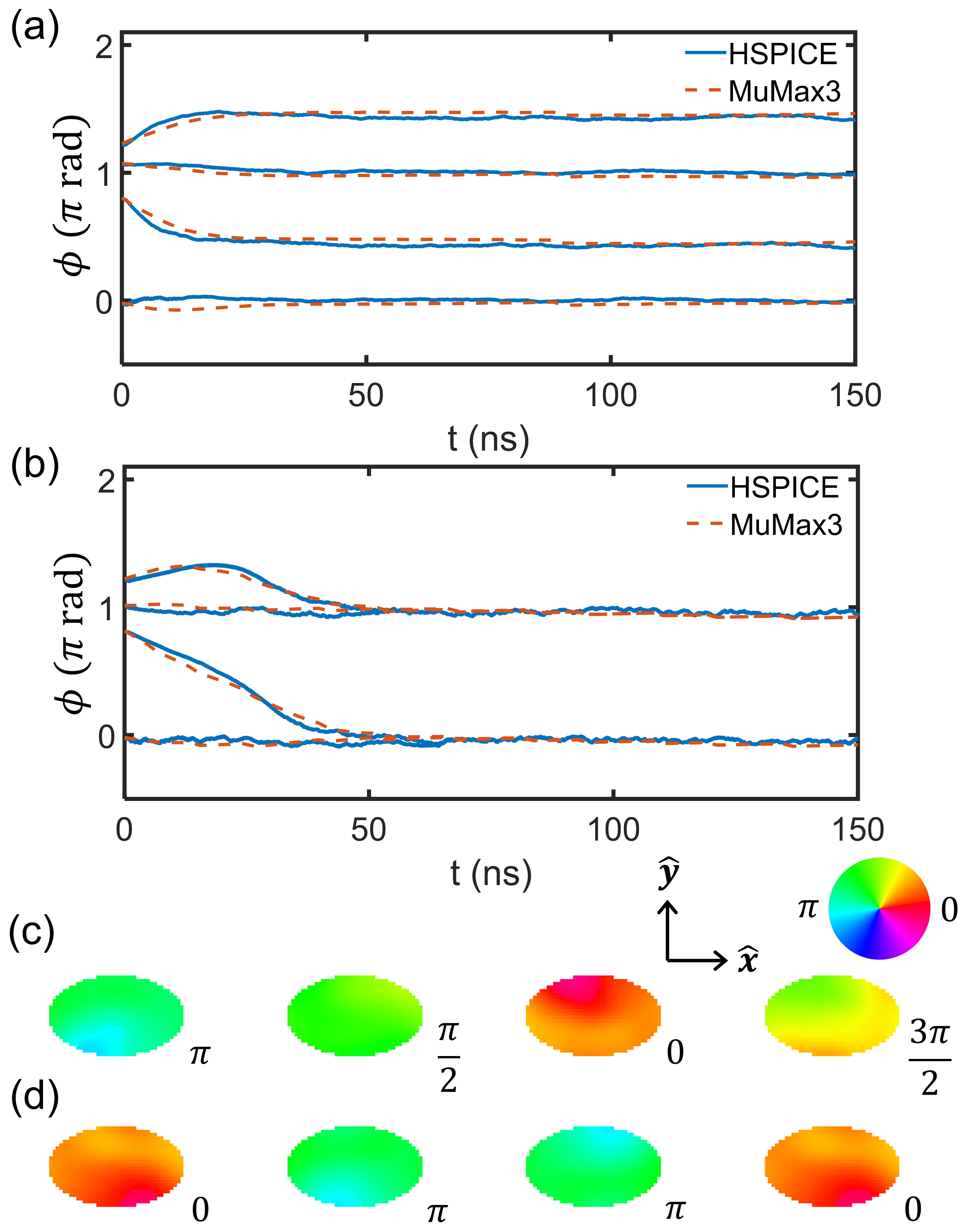}
 \caption{\label{fig:hspice_mumax_phase_comparison}Phase dynamics of 4-node all-to-all coupled network solving the Ising model with $J_{ij}=-1$ in HSPICE and MuMax3 (neglecting thermal noise) (a) with first harmonic coupling only and (b) with addition of binarizing second harmonic signal. Without second harmonic locking (a), the phases spread across $0-2\pi$ and the ground state is not clearly visible. Once a second harmonic signal is applied (b), the phases clearly binarize and settle to the ground state $\{\phi\}=\{0,0,\pi,\pi\}$. (c,d) In-plane spatial plots of the free layer magnetization in MuMax3 corresponding to the final phase configuration at $t=150$ ns in (a,b) respectively. Inter-oscillator spacing has been reduced for visibility.}
\end{figure}

Using the abstract SHNO device model, we can directly construct a coupled oscillator network and simulate its performance at the circuit level. One branch of the electrical coupling circuit simulated in HSPICE is shown in Fig. \ref{fig:simp_circ_diagram}. A buffer on the oscillator output is used to match the SHNO impedance with the following stage and an rf gain block ensures an adequate coupling signal amplitude. We use Texas Instruments LMH5401 and Texas Instruments TRF37D73 for these two devices. The buffer has a fully differential architecture that provides oscillator outputs of both positive and negative polarities to the coupling circuit, allowing problems with any coupling coefficient signs to be modeled. The coupling conductances are set by variable resistors, which can be realized using voltage-programmable memristors. The first harmonic coupling signal to the input of node $i$ is given by
\begin{equation}
    V_{1f}\sin(\omega_g t+\phi_{e1})=A_{v}\sum_{j=1}^N |V_{out,j}|\frac{R_{HM}}{R_{ij}}\sin(\omega_g t+\phi_j)
\end{equation}
where $A_{v}=5.5$ is the net signal loop gain, $|V_{out,j}|$ is the amplitude of the output signal from oscillator $j$ in Eq. (\ref{eq:vout}) (with DC component filtered), and $R_{ij}$ is the resistance linking oscillators $i$ and $j$. Substituting into Eq. (\ref{eq:isf_adler}), the Ising model coefficients in this coupling scheme will be proportional to the tunable internode conductances $1/R_{ij}$. Finally, in a realistic circuit, delay lines need to be designed such that any finite parasitic phase shift around the coupling loop is compensated, avoiding associated time-delay coupling complications \cite{tiberkevichSensitivityExternalSignals2014,tsunegiSelfInjectionLockingVortex2016,tiberkevichPhaselockingFrustrationArray2009,khalsaCriticalCurrentLinewidth2015}.

To obtain a baseline result for the coupling behavior, we simulate 4 all-to-all coupled oscillators (Fig. \ref{fig:max_cut_example}) in HSPICE with uniform coupling $J_{ij}=-1$ where the minimum-energy (ground state) phase configuration corresponds to any permutation of $\{\phi\}=\{0,0,\pi,\pi\}$. Meanwhile, to verify the validity of the circuit simulations, we replicate the same problem in micromagnetic simulations by modifying the open-source MuMax3 code to support electrical signal generation and coupling as described in Appendix \ref{app:Electrically Coupled Oscillators}. The simulated phase dynamics from these two parallel approaches are shown in Figs. \ref{fig:hspice_mumax_phase_comparison}(a) and \ref{fig:hspice_mumax_phase_comparison}(b). In Fig. \ref{fig:hspice_mumax_phase_comparison}(a), only the first harmonic injection locking signal is introduced to each node, while in Fig. \ref{fig:hspice_mumax_phase_comparison}(b), a global second harmonic signal is included with amplitude $V_{2f}=40$ mV. With the second harmonic signal, the phases of the SHNOs binarize and arrive at the ground state solution. Spatial plots of the magnetization of the unbinarized and binarized oscillators from micromagnetic simulations are also shown in Figs. \ref{fig:hspice_mumax_phase_comparison}(c) and \ref{fig:hspice_mumax_phase_comparison}(d). The key features of the phase dynamics, including time-to-solution and steady-state configuration, are very well matched between micromagnetic and circuit simulations. Slight discrepancies in the shape of the phase curves can be attributed to the more abstract macrospin-based modeling used in HSPICE. Nevertheless, the circuit model accurately reproduces the phase dynamics of the coupled oscillator network from micromagnetic simulations and can be simulated in less than 1\% of the time.

\section{\label{sec:Performance of SHNO Ising Machine}Statistical Performance}

\begin{figure*}
 \includegraphics[width=17.2cm]{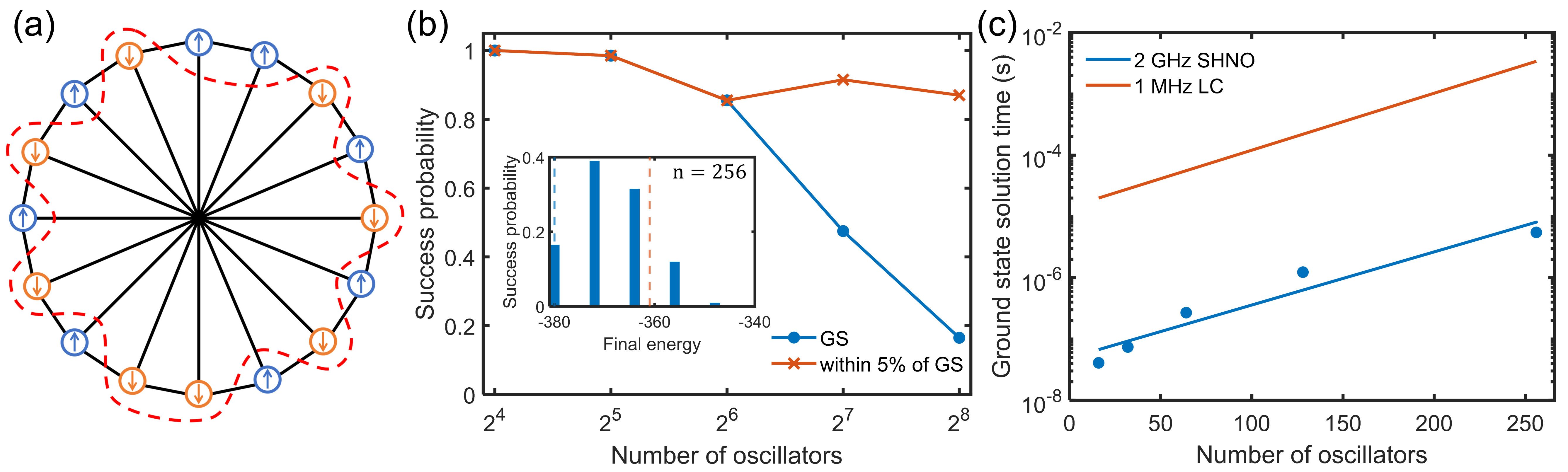}
 \caption{\label{fig:n_osc_figure}(a) 16-node M\"{o}bius ladder graph architecture used in Section \ref{sec:Performance of SHNO Ising Machine}. The MAX-CUT solution for an unweighted graph ($w_{ij}=1$) is shown by the dotted line partitioning the graph into two opposite-spin subgraphs. (b) Probability of reaching GS and within 5\% of GS vs. $n$. Inset: Histogram of final energy states for $n=256$ (blue dotted line is $E_{GS}=-380$, orange dotted line is within 5\% of GS). (c) Solution time of SHNO Ising machine (blue scatters) as a function of $n$. The blue line shows the trend. The orange line represents the expected solution time of a 1 MHz LC oscillator Ising machine reproduced from Chou \textit{et al.} \cite{chouAnalogCoupledOscillator2019}. \footnote{Licensed under CC BY 4.0 (https://creativecommons.org/licenses/by/4.0/)}}
\end{figure*}

\begin{table*}
\caption{\label{tab:performance_comparison}
Performance comparison between the SHNO Ising machine in this work and existing technologies. Time and energy values are standardized to a size-100 problem. Acronym key: D-WAVE = quantum annealing, OPO = optical parametric oscillator, PTNO = phase transition nano-oscillator.}
\begin{ruledtabular}
\begin{tabular}{l l l l l l l l}
& CPU & GPU & D-WAVE & OPO & LC & PTNO & SHNO \\
& \cite{duttaIsingHamiltonianSolver2020} & \cite{kingEmulatingCoherentIsing2018} & \cite{boixoEvidenceQuantumAnnealing2014} & \cite{inagakiCoherentIsingMachine2016,mcmahonFullyProgrammable100spin} & \cite{chouAnalogCoupledOscillator2019,chiccoAnalysisPowerConsumption2017} & \cite{duttaIsingHamiltonianSolver2020,duttaProgrammableCoupledOscillators2019} & [this work] \\
\hline
Solution time & 246ms & 100$\mu$s & 30ms & 2ms & 2.5ms & 25.5$\mu$s & 359ns \\
Power & 60W & $\leq$250W & 25kW & - & 250mW & 1.2mW & 48.6mW \\
Energy to solution & 14.8J & $\leq$25mJ & 750J & - & 625$\mu$J & 30.6nJ & 17.5nJ \\
Energy efficiency (sol/s/W) & 6.7$\times10^{-2}$ & $\geq$40 & 1.3$\times10^{-3}$ & - & 1.6$\times10^3$ & 3.26$\times10^7$ & 5.73$\times10^{7}$ \\
Size (one element) & - & - & - & 1km fiber & 0.1mm$^2$ & 0.2$\mu$m$^2$ & 0.024$\mu$m$^2$ \\
Room temperature & Y & Y & N & Y & Y & Y & Y \\
\end{tabular}
\end{ruledtabular}
\end{table*}

Finally, we scale the HSPICE circuit model to larger coupled oscillator networks to study the performance of the SHNO Ising machine. The graph topology we use is an undirected ($J_{ij}=J_{ji}$) M\"{o}bius ladder graph, as shown in Fig. \ref{fig:n_osc_figure}(a) for the 16-node case. We focus on a common set of problems called MAX-CUT that involve partitioning a weighted graph into two subgraphs such that the sum of the edge weights $w_{ij}$ linking nodes in the two subgraphs is maximized \cite{karpComputationalComplexityCombinatorial1975,yannakakisNodeandEdgedeletionNPcomplete1978,kalinincomplexity2020}. In the example shown in Fig. \ref{fig:n_osc_figure}(a), this is to find a cut line (dotted curve) that maximizes $\sum_{\mathrm{cut}} w_{ij}$, where $w_{ij}$ correspond to the edges connecting nodes from different subgraphs. MAX-CUT problems are mapped to the Ising Hamiltonian in Eq. (\ref{eq:ising_model}) by coupling coefficients $J_{ij}=-w_{ij}$ and spin values $s_i$ representing the subgraph a node belongs to. When mapping between the edge weights $w_{ij}$ and coupling conductances $1/R_{ij}$, we scale by a constant factor such that the coupling signal amplitude is bounded by $V_{1f}\leq80$ mV.

First, we demonstrate the scaling of the solution time and probability with the number of coupled oscillators $n$. We use unweighted ($w_{ij}=1$) M\"{o}bius ladder graphs of sizes $n=16-256$. Because the phase binarization process is slow for large arrays, an annealing schedule is employed where $V_{2f}$ is stepped between 0 and $50$ mV every 100 ns, resulting in strong phase binarization \cite{wangOIMOscillatorBasedIsing2019}. Graphs of this size are already difficult to solve by brute force calculation, but in this case the MAX-CUT solution always takes the form shown in Fig. \ref{fig:n_osc_figure}(a) where the ground state energy can be generalized as $E_{GS}(n)=-\frac{3}{2}n+4$ \cite{kalinincomplexity2020}. In Fig. \ref{fig:n_osc_figure}(b), we show the probability of reaching GS and a close-to-optimal solution (within 5\% of GS) as a function of $n$. Due to simulation time overhead, thermal noise is not considered. The data is aggregated over 1000 simulations with randomly generated initial phase configurations. Though the GS probability falls to 17\% by $n=256$, the phases settle within 5\% of GS with over 85\% probability even up to the largest simulated array. The inset to Fig. \ref{fig:n_osc_figure}(b) shows a histogram of the final energy state distribution for $n=256$. We see that the final phase states are consistently distributed among a few local energy minima very close to GS, representing close-to-optimal solutions.

In Fig. \ref{fig:n_osc_figure}(c), we show the solution time $t_{sol}$ as a function of $n$, with $t_{sol}=t_{GS}/P_{GS}$ where $t_{GS}$ is the solution time of runs that reach GS and $P_{GS}$ is the GS probability from Fig. \ref{fig:n_osc_figure}(b) \cite{chouAnalogCoupledOscillator2019}. Compared to a 1 MHz LC oscillator Ising machine \cite{chouAnalogCoupledOscillator2019}, the GHz SHNO Ising machine achieves solution times 3 orders of magnitude lower due to the much higher operating frequency. The fast solution time is arguably the most significant benefit of the SHNO Ising machine, as it enables solving larger and more difficult problems on reasonable timescales and with less energy. In Table \ref{tab:performance_comparison}, we compare the solution time and energy consumption of the SHNO Ising machine against existing technologies. We neglect the power consumption of off-the-shelf amplifiers in our implementation. By using integrated on-chip amplifier circuits rather than off-the-shelf components to reduce the active power \cite{chenIntegrationGMRbasedSpin2015}, we can potentially realize a significantly more energy-efficient Ising machine than previous implementations.

\begin{figure}
 \centering
 \includegraphics[width=8cm]{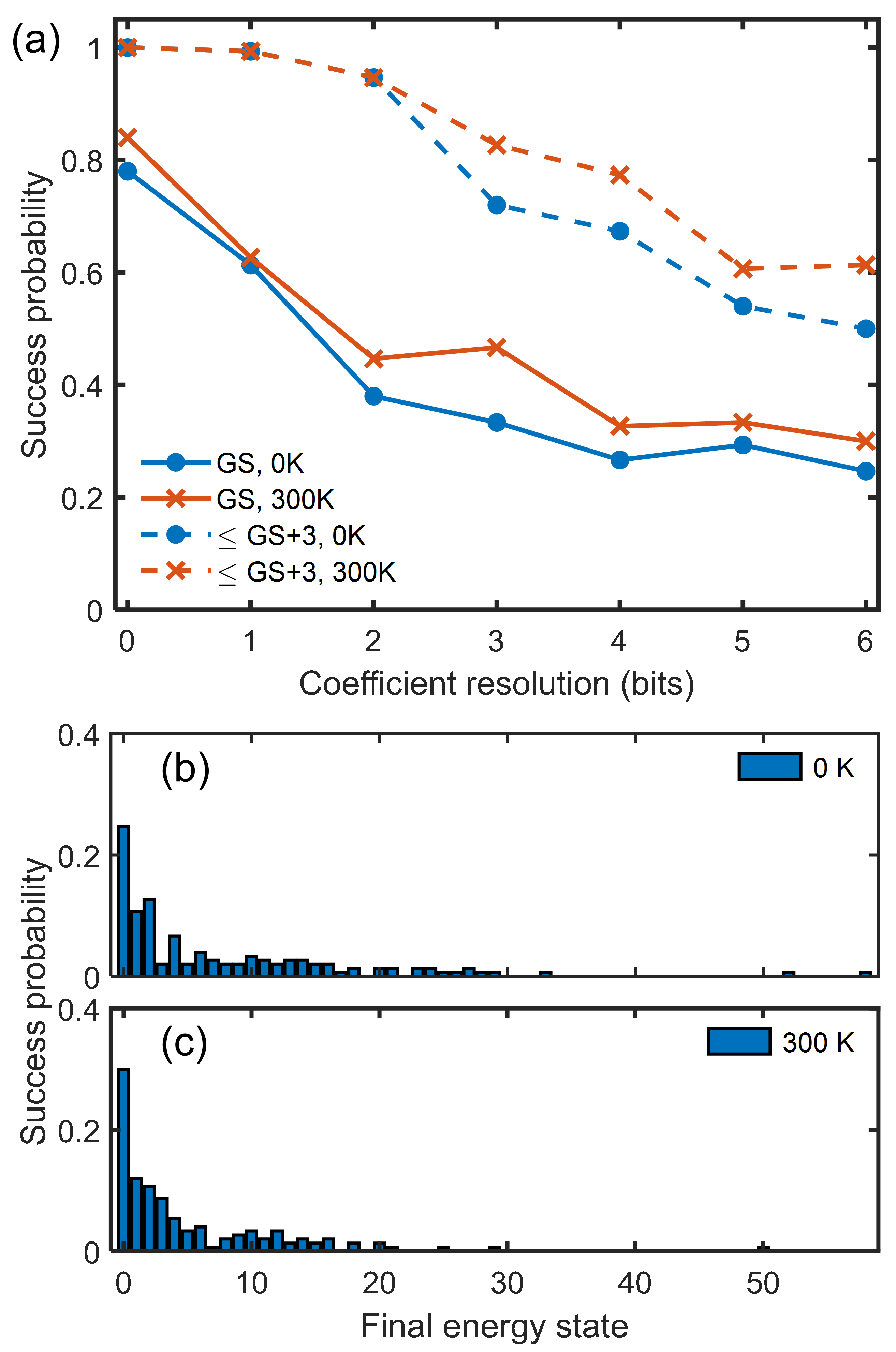}
 \caption{\label{fig:16_node_coeff} (a) Probability of reaching the Ising Hamiltonian ground state (GS) and within 3 states of GS ($\leq$GS+3) vs. coupling coefficient bit resolution with and without thermal noise. (b,c) Histograms of final energy states for problems with 6-bit coefficients (b) without thermal noise and (c) with thermal noise. The ground state is 0, so lower-index states represent closer-to-optimal solutions.}
\end{figure}

Next, we study 16-node MAX-CUT problems with randomly generated coupling coefficients and thermal noise. We vary the bit resolution of the coupling coefficients $J_{ij}$ from 0-6 bits, where higher-resolution problems are generally harder to solve due to having more complex energy landscapes with shallower energy minima \cite{chouAnalogCoupledOscillator2019}. 0-bit resolution represents constant-magnitude coefficients $|J_{ij}|=1$, while 6-bit coefficients take any integer value $|J_{ij}|\leq 2^6$. In all cases, the sign of $J_{ij}$ is randomly assigned between each pair of coupled nodes. A constant second harmonic signal with amplitude $V_{2f}=30$ mV is applied. For this relatively small scale problem, all the energy levels of the Ising Hamiltonian with all possible node states $\{s_i\}$ can be enumerated using a brute force approach, providing a standard solution to compare to. In Fig. \ref{fig:16_node_coeff}(a), we show the probability of reaching GS or within three states of GS ($\leq$GS+3) on an identical set of benchmark problems with and without thermal noise. Each set of results is aggregated over 1000 simulation runs with randomly generated coefficients and initial oscillator phases. Across all bit resolutions, thermal noise consistently increases the probability of reaching the ground state or another low-energy state, representing optimal and close-to-optimal solutions of the MAX-CUT problem. Histograms of the final energy states for 6-bit coefficients with and without thermal noise in Figs. \ref{fig:16_node_coeff}(b) and \ref{fig:16_node_coeff}(c) provide more insight into this trend. The presence of noise enables the oscillator phases to escape local energy minima and settle closer to the ground state, forming a higher concentration of low-energy states. The total number of non-degenerate phase states in this case is $2^{15}$, whereas the highest state observed in Figs. \ref{fig:16_node_coeff}(b) and \ref{fig:16_node_coeff}(c) is 58, indicating that the coupled oscillators reach very close-to-optimal solutions in all trials.

\section{Conclusion}

In this paper, we have presented a detailed analytical study of the electrically coupled SHNO Ising machine. We began by developing an ISF-based analytical model for the injection locking strength of a SHNO that can be used to model oscillators with nontrivial precession trajectories. By explicitly including the shape of the precession trajectory in our model, we could accurately predict injection locking not only at the oscillator's fundamental frequency but also at higher harmonic frequencies. We then integrated this model into a Verilog-A device for efficient circuit-level modeling of the Ising machine. Through scaling simulations, we demonstrated that the GHz-frequency SHNO array can solve combinatorial optimization problems on nanosecond timescales, achieving orders of magnitude improvements in solution speed and energy efficiency compared to previously proposed Ising machines. Thermal phase noise is also expected to increase the probability of reaching the ground state or other low-energy states, allowing the SHNO Ising machine to be operated at room temperature. Our results provide useful quantitative insights and tools to enable the realization of a high-speed, energy-efficient, and ultra-scalable Ising machine employing SHNOs.

\appendix

\section{\label{app:Single Oscillator}Micromagnetic Simulation of Single Oscillator}

For our micromagnetic simulations, we use the GPU-accelerated, open-source MuMax3 code \cite{vansteenkisteDesignVerificationMuMax32014}. The simulated SHNO has the general structure shown in Fig. \ref{fig:mtj_axis_def}(a). The in-plane dimensions of the MTJ are $200\times120$ nm$^2$ ($x\times y$) with a magnetic field $B_{ext}=20$ mT applied along the short axis of the MTJ. Only the CoFeB free layer with thickness $t_{FL}=$1.5 nm is explicitly modeled with spin injection parameters assuming a Pt heavy metal layer with $\theta_{SH}=0.07$. The CoFeB material parameters are as follows: saturation magnetization $M_s=800$ kA/m, exchange stiffness $A_{ex}=13$ pJ/m, Gilbert damping $\alpha=0.02$, perpendicular magnetic anisotropy constant $K_{u}=180$ kJ/m$^3$. Because the injected spins are polarized in-plane, a larger out-of-plane demagnetizing field (smaller PMA) should increase the oscillator phase sensitivity to spin torque based on Eq. (\ref{eq:isf_rf_fourier}), so we choose to reduce the out-of-plane demagnetizing field by only half. The free layer is discretized into cells with dimensions $5\times 5\times 1.5$ nm$^3$ in $x$, $y$, and $z$ respectively. The threshold DC voltage to observe auto-oscillations is $V_c=60$ mV ($J_c=V_c/(R_{HM}w_{HM}t_{HM})=3.0\times10^{11}$ A/m$^2$), and the DC operating point $V_{DC}=88$ mV ($J_{DC}=4.4\times10^{11}$ A/m$^2$) is chosen for the large-angle oscillation in the $x$-$y$ plane. In micromagnetic simulations, the temperature is set to 0 K.

\begin{figure}
    \centering
    \includegraphics[width=7.5cm]{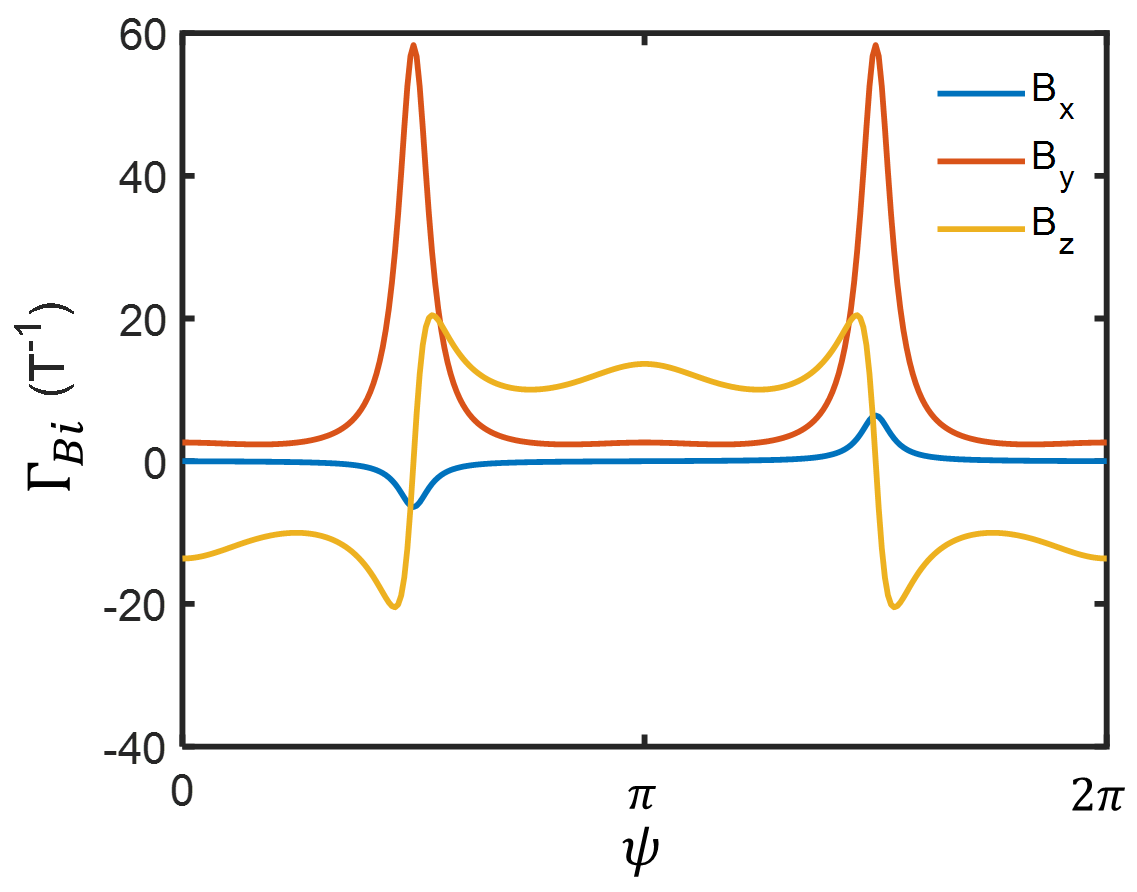}
    \caption{ISFs for perturbation by magnetic fields along $x,y,z$ axes (Eqs. (\ref{eq:isf_bx})-(\ref{eq:isf_bz})) based on SHNO with clamshell orbit.}
    \label{fig:field_isf}
\end{figure}

\section{Magnetic Field Impulse Sensitivity Function} \label{sec:Magnetic Field Impulse Sensitivity Function}

The derivation of the ISF for magnetic field perturbation follows from a similar analysis as for the spin torque ISF in Section \ref{sec:Impulse Sensitivity Function}. In this case, the general ISF expression is
\begin{equation}
    \Gamma_{Bi}(t)=\frac{\Delta\boldsymbol{\tau}_{Bi}}{B}\cdot\frac{\dot{\mathbf{m}}_{eq}}{|\dot{\mathbf{m}}_{eq}|^2}
    \label{eq:shno_isf_bi}
\end{equation}
where $i=\{x,y,z\}$ and $B$ is the magnitude of the perturbing field. The relevant torque terms from Eq. (\ref{eq:llgs}) are the first and second terms; however, the second term is negligible due to low magnetic damping. The torque due to a perturbing magnetic field along $i$ direction (considered separate from the DC effective field $\mathbf{B}_{eff}$) is then
\begin{equation}
    \Delta\boldsymbol{\tau}_{Bi}=-\gamma_e(\mathbf{m}\times\mathbf{B}_i)
\end{equation}
The ISFs can again be represented by Fourier series
\begin{align}
\Gamma_{Bx}(t)&=\frac{\omega_g\gamma_e}{2|\dot{\mathbf{m}}_{eq}|^2}\sum_{n=\mathrm{odd}}b''_n\frac{1}{n^2}\sin(n\psi) \label{eq:isf_bx}\\
\Gamma_{By}(t)&=\frac{\omega_g\gamma_e}{|\dot{\mathbf{m}}_{eq}|^2}m_x m_z \label{eq:isf_by}\\
\Gamma_{Bz}(t)&=\frac{\omega_g\gamma_e}{2|\dot{\mathbf{m}}_{eq}|^2}\sum_{n=\mathrm{odd}}a''_n\frac{1}{n^2}\cos(n\psi) \label{eq:isf_bz}
\end{align}
with non-zero coefficients
\begin{align*}
b''_1&=(3m_y-2m_{y0})m_z \\
b''_3&=m_y m_z \\
a''_1&=-(3m_y+2m_{y0})m_x \\
a''_3&=m_x m_y
\end{align*}
The ISFs in Eqs. (\ref{eq:isf_bx})-(\ref{eq:isf_bz}) are plotted in Fig. \ref{fig:field_isf}. We see that the highest sensitivity to external field perturbations occurs at $\psi=\pi/2$ and $3\pi/2$, where the conversion of thermal fluctuations (modeled as an effective field in Eq. (\ref{eq:b_therm})) to phase noise will be greatest. While it is not a focus of this paper, we note that one may use the ISF as a tool to analytically predict and design the phase noise performance of the oscillator \cite{hajimiriGeneralTheoryPhase1998,demirPhaseNoiseOscillators2000}.

\section{\label{app:Electrically Coupled Oscillators}Micromagnetic Simulation of Electrically Coupled Oscillators}

The MuMax3 code is modified to enable simulation of electrical signal generation and coupling of up to 4 SHNOs \footnote{Modified MuMax3 source code. \url{https://github.com/bcmcgold/3}.}. The oscillators are simulated within a single strip geometry with dimensions $6000\times w_{MTJ} \times t_{FL}$ nm$^3$ and are each spaced apart by 2 $\mu$m to minimize coupling via the stray fields. The devices used in the coupling simulations have the same parameters as the single device in Appendix \ref{app:Single Oscillator} and the electrical parameters are the same as in HSPICE simulations (Table \ref{tab:va_parameters}).

To construct the electrical output signal of each oscillator, a new magnetization-dependent quantity describing the oscillator magnetoresistance is defined in MuMax3. The oscillating MTJ resistance assuming a fixed layer polarized along the $+x$ axis is
\begin{equation}
    R_{out}(t)=\frac{R_P-R_{AP}}{2}m_x(t)+\frac{R_P+R_{AP}}{2}
    \label{eq:Rout}
\end{equation}
Next, an MTJ read current density $J_{MTJ}$ is specified. Gain elements are not modeled so $J_{MTJ}$ implicitly contains any necessary gain factor for the coupling signal. Finally, a 6-element coupling conductance vector $G_{ij}=[G_{11}, G_{12}, G_{13}, G_{23}, G_{24}, G_{34}]$ denoting the conductances linking oscillators $i$ and $j$ is specified to fully describe an all-to-all connected 4-node network (Fig. \ref{fig:max_cut_example}).

A coupling current density is applied to SHNO $i$ by adding the following expression to the existing spin transfer torque term in MuMax3
\begin{equation}
    J_{cpl,i}=J_{MTJ}\sum_{j,i\neq j}^4 G_{ij}\frac{R_P-R_{AP}}{2}m_{x,j}(t)
    \label{eq:Jcpl}
\end{equation}
Modified versions of the RK45 and RK56 numerical solvers are written with Eq. (\ref{eq:Jcpl}) appended to the MuMax3 equation set solved for each oscillator \cite{vansteenkisteDesignVerificationMuMax32014}.

%

\end{document}